# Excitonic shift current induced broadband THz pulse emission efficiency of layered single crystalline $MoS_2$

Neetesh Dhakar and Sunil Kumar*

*Femtosecond Spectroscopy and Nonlinear Photonics Laboratory, Department of Physics,*

*Indian Institute of Technology Delhi, Hauz Khas, New Delhi 110016, India*

*Corresponding author: kumarsunil@physics.iitd.ac.in

**Following the ultrafast photoexcitation of a semiconductor, it embodies competing dynamics among photocarriers, many-body transient states of highly energetic excitons, and electron-hole liquid. Here, we show that femtosecond optical pulse excitation induces transient excitonic shift current contributing to stronger THz emission from a single crystalline bulk $MoS_2$ at low temperatures. The control of dominating excitonic shift current is elucidated from excitation density dependent experiments at varying temperatures. A strong decrease in the excitonic contribution beyond a critical fluence of 150μJ/cm$^2$ is observed at a very low temperature of 20K. This behavior suggests the formation of a new quantum condensate, i.e., the electron-hole liquid, in the regime when the exciton density is overwhelmingly large that the average spacing between exciton pairs is comparable to the exciton radius. Furthermore, the exciton density dependent THz emission at varying temperatures is consistent with the Varshni model and the crystal Debye temperature of 260K.**

The charge-neutral bound electron-hole pairs, namely, excitons in semiconductors, are held together by the Coulombic force between them, where the exciton binding energy varies depending on the dielectric constant of the semiconductor and the effective mass of the carriers [1, 2]. In most semiconductors, the excitons are highly unstable at room temperature. In the van der Waals layered transition metal dichalcogenides (TMDs), upon appropriate optical excitation, much long-lived and stable excitons are formed quickly as one lowers the bulk sample temperature from the room temperature value[3-5]. In fact, exciton binding energies in thin TMDs are reasonably high to have stable excitons even at room temperature. By means of ultrashort laser pulses, the formation and decay dynamics of the excitons can be studied. Molybdenum disulfide ($MoS_2$) is one of the most popular TMD material systems. By ultrafast optical excitation, the transition of the bulk $MoS_2$ from an exciton-rich state at the low excitation fluences to one at different stages of macroscopic transient states of the correlated electron-hole pairs, such as the electron-hole plasma and electron-hole liquid (EHL), depending on the excitation fluences used, are known to us to some extent through transient photoluminescence and transient absorption studies[6-8].

An ultrafast photoexcitation of a semiconducting bulk $MoS_2$ leads to the creation of various transient photocurrents in the materials, either directly by the generation and decay of the photoexcited carrier density or the excitons. A dynamic coherent exciton population, generated directly by ultrafast optical excitation, can contribute to the formation of transient photocurrent, namely, the exciton shift-current[9-11]. Such induced photocurrents can lead to the emission of electromagnetic radiation from the material, that in the terahertz (THz) realm[12-14], if photoexcited by femtosecond laser pulses. In $MoS_2$, an azimuth-dependent THz emission behavior has been associated to nonlinear optical contributions rather than just the induced transient photoconductivity due to real photocarriers[12, 15]. For the above bandgap ultrafast photoexcitation, THz radiation emission from bulk $MoS_2$ was considered to be originating primarily due to the surface depletion field mediated transient photocurrent[13, 16]. An emergence of excitonic shift current, for the case, when the exciton absorption is well separated from the conduction continuum, and the ultrafast excitation is performed at low temperatures so as the thermal annihilation of excitons into free carriers is minimized[17], it is possible to realize it contributing to the THz radiation emission. For example, the ultrafast laser-induced exciton shift current in cadmium sulfide was studied for free space THz generation was reported recently[17]. Hu *et al.*[21], employed the ultrafast electron diffraction method to explore the many-body structural dynamics triggered by near-resonant excitation of low-energy indirect excitons in $MoS_2$. It is anticipated that the exciton-rich semiconducting bulk $MoS_2$ crystal must also garner the above-mentioned transient charge and excitonic shift currents, thereby making them relevant even at the THz frequencies over and above the well-tested other optoelectronic applications[18-20].

In the current work, we have determined pulsed THz emission properties of bulk single-crystalline $MoS_2$. A continuous enhancement in the THz generation efficiency at low temperatures, an overall THz generation efficiency getting more than doubled when the crystal temperature lowers from room temperature to around $20\,K$, is observed. Supported by excitation density dependent measurements, the above behavior has been attributed to originating from an ultrafast excitonic shift current due to direct excitation of low-energy indirect excitons by broadband femtosecond laser pulses centered at 800 nm. The results have been analyzed to quantitatively extract other contributors to the THz emission, i.e., the optical rectification and surface depletion field induced photocurrent, along with the dominating excitonic shift current, the latter being stipulated by the just below direct bandgap but above the indirect bandgap photoexcitation of the crystal.

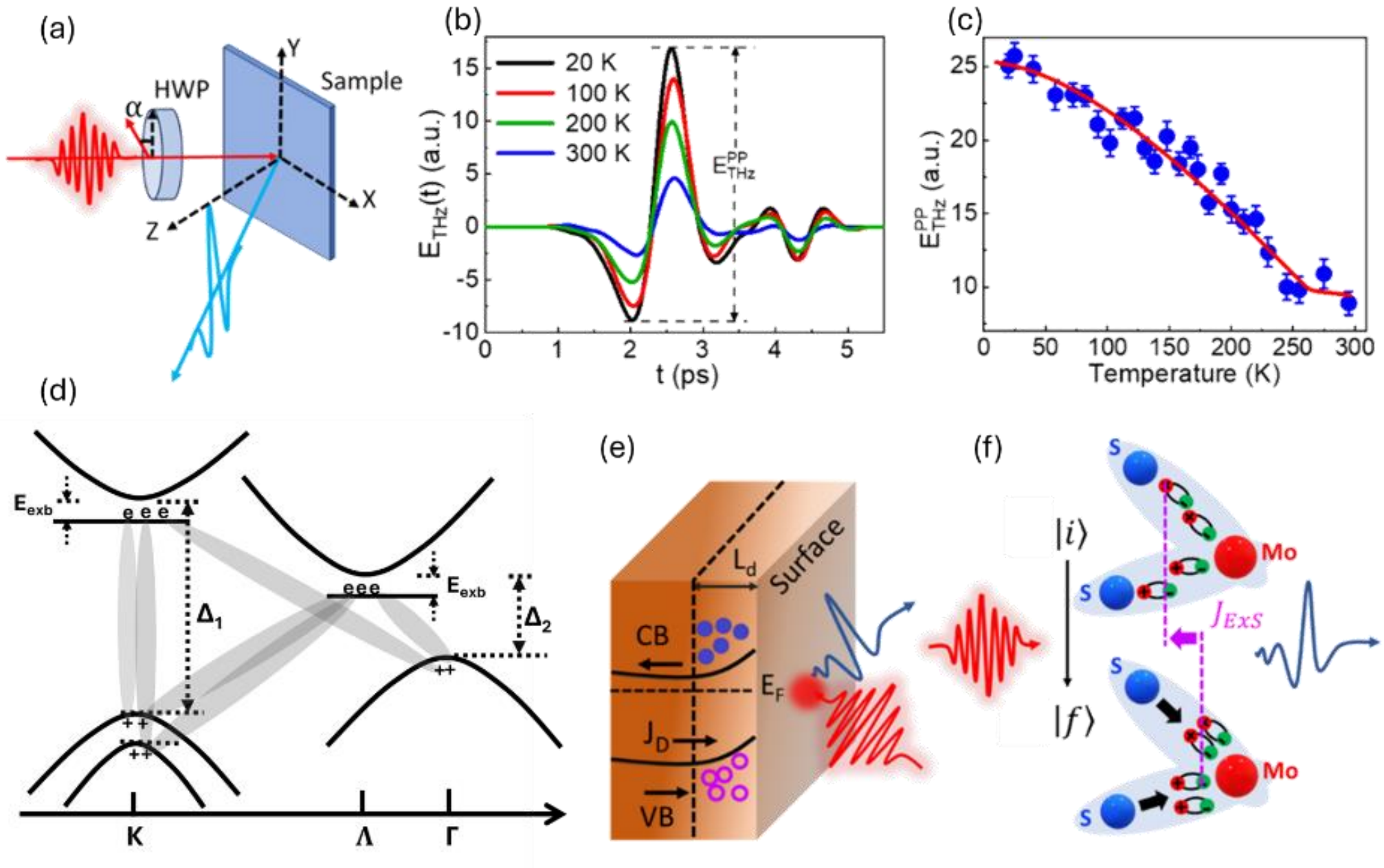

**Figure 1. (a) Schematic for sample excitation by broadband femtosecond laser pulses of variable linear polarization and THz pulse emission in the reflection configuration. HWP: broadband zero-order half-wave plate at a variable angle α. (b) Typical THz pulses emitted from $MoS_2$ crystal shown for four temperatures, 300 $K$, 200 $K$, 100 $K$, and 20 $K$ with a pump excitation fluence fixed at 120 $\mu J/cm^2$. The peak-to-peak value of the THz field, $E_{THz}^{PP}$ is indicated. (c) Temperature dependence of $E_{THz}^{PP}$ representing a monotonous increase in the THz generation efficiency with the decreasing temperature. The solid curve is fit to the data (see text). (d) Schematic electronic band-diagram and various excitons indicated by oval shapes. The $\Gamma_v K_c$ exciton between the Γ-point in the valence band and the K-point near the conduction band and another, $\Gamma_v \Lambda_c$ exciton between the Γ-point in the valence-band and the Λ-point near the conduction band are relevant for excitation pulses with central photon energy, $E_p = 1.5\ eV$. Schematics for the THz pulse generation by (e) surface depletion field-mediated transient drift current $J_D$ and (f) ultrafast excitonic shift current $J_{ExS}$.**

A schematic is shown in Fig. 1(a) depicting the photoexcitation of the $MoS_2$ crystal by femtosecond laser pulses of variable polarization and emission of THz pulses in the reflection geometry. Further details about the experimental setup are provided in the Supplementary Information. Femtosecond laser pulses having a pulse duration of ~35 $fs$, central wavelength (photon energy) $\lambda_p$ $(E_p)$~$800\ nm$ $(1.55\ eV)$ and pulse repetition rate of 1 kHz, were taken from a Ti:Sapphire regenerative amplifier for the ultrafast photoexcitation of the sample by a collimated beam (diameter of ~$3\ mm$) as well as the detection of emitted THz pulses. The THz pulse detection for a specific polarization was achieved by using electro-optic sampling in a nonlinear crystal in our home-built setup[22-24]. The pump excitation fluence in the experiments was controlled by using various neutral density filters, while the temperature (T) of the sample was varied by placing it inside a closed-cycle liquid helium optical cryostat.

For a fixed excitation polarization and fluence of ~$120\ \mu J/cm^2$, the amount of the generated THz radiation continuously increased as the sample temperature was lowered from room temperature. At a few representative temperatures, the real-time variation of the electric field of the radiated THz pulses as obtained from these experiments is presented in Fig. 1(b). The peak-to-peak value, $E_{THz}^{PP}$, of the corresponding THz field as indicated in Fig. 1(b), is a good measure of the THz generation efficiency under the given conditions. Hence, the same has been used to plot the outcomes for the THz generation efficiency under fixed excitation conditions but in the entire range of the sample temperatures in Fig. 1(c).

At low temperatures, $MoS_2$ is exciton-rich, and hence those must contribute to THz emission for the observed temperature dependence in Fig. 1(c). A schematic electronic band diagram is shown in Fig. 1(d) with $\Delta_1 = 1.85\ eV$ and $\Delta_2 = 1.3\ eV$ as the direct and indirect electronic bandgap energies at the K- and Γ-points, respectively, in the bulk $MoS_2$ crystal. Various possibilities are indicated by oval shapes for the exciton formation between various high symmetry points in the valence (v) and conduction (c) bands of an optically excited crystal. The exciton binding energy is represented by $E_{exb}$~ $50\ meV$. In addition to the direct excitons ($E_A = 1.8\ eV$ as $K_{v1}K_c$ and $E_B = 2\ eV$ as $K_{v2}K_c$), we have four other possibilities for the indirect excitons as shown. Among them, the $\Gamma_v K_c$ and $\Gamma_v \Lambda_c$ excitons having energy in the range of 1.33-1.59 eV and 1.23-1.44 eV, respectively,[21, 25] are directly relevant for the current observations. The central photon energy of the femtosecond pulses used here is nearly in resonance with $\Gamma_v K_c$ and in close proximity of $\Gamma_v \Lambda_c$ indirect excitons.[21] Since, we have utilized here a bulk $MoS_2$ crystal in its 3R phase (see S3 in the Supplementary information), which lacks inversion symmetry,[32] hence it can exhibit the generation of an ultrafast excitonic shift current. In the following, we describe the contributions of a dominating transient excitonic shift current $J_{ExS}$ at low temperatures and dominating transient drift photocurrent $J_D$ due to photocarriers at room temperature towards the THz generation from $MoS_2$.

The photogenerated excitons are less stable or short-lived at room temperature due to the presence of active phonons and dissociate either to free carriers or annihilate by recombination. As such, significant phonon scattering reduces the exciton binding energy and stability, leading to a diminishing coherent exciton population and hence the respective contribution to THz emission. Under these conditions, THz emission is predominantly governed by free carriers through the transient photocurrent effect (TPE), in which photogenerated carriers are accelerated by internal electric fields. The internal electric field can arise due to two reasons, i.e., the surface depletion field and the photo Dember field. The surface depletion field at the air-material interface of any semiconductor is like what is depicted in Fig. 1(e) under which the photocarriers drift, causing a transient drift photocurrent $J_D$. It is given as $J_D = ne\mu E_s$, $n$ being the photocarrier density, $e$ the electronic charge, $\mu$ the carrier mobility, and $E_s$ the surface depletion field. For the Dember field to cause a photocurrent, the semiconductor necessitates to have a significant difference in the electron and hole mobilities. This is the reason for it to have been attributed in the THz emission from narrow bandgap semiconductors, such as InAs and InSb.[26] In bulk $MoS_2$, the electron and hole mobilities are quite comparable[27], and hence THz emission due to the photo-Dember effect can be safely ignored. THz emission from other effects such as optical rectification (OR), photogenerated free carrier shift current and photon drag effect (PDE) also needs to be considered. The contribution from the shift current is minimal in our case due to below bandgap excitation. For the case of PDE in which, photocarrier acceleration takes place due to direct transfer of incident photon momentum,[28] THz emission for oblique incidence has been predominantly observed in materials possessing high concentrations of free carriers[29], such as metals, heavily doped semiconductors, graphene, etc. In bulk $MoS_2$, the intrinsic carrier concentration is low, and in fact, the carrier mobilities are also quite low at all temperatures, at least by two orders of magnitude lower when compared with graphene, a case in which a significant effect is seen.[30] Therefore, THz emission by PDE can also be ignored in our case here. On the other hand, OR at the surface and in the bulk of the noncentrosymmetric crystal is always present. It is considered to relatively weakly contribute to the THz emission across all temperatures.[31]

At low temperatures, the excitons in $MoS_2$ become much more stable and their lifetime also increase due to the limited involvement of phonons and helps in forming an additional photocurrent, i.e., the exciton shift current $J_{ExS}$. Our observation that the overall THz emission from $MoS_2$ gets nearly doubled at low temperatures clearly signifies the role of the $J_{ExS}$. A possible mechanism of the transient $J_{ExS}$ generation[9] accompanying the usual exciton formation at below bandgap photoexcitation at low temperatures, is schematically shown in Fig. 1(f). The excitons transition between $|i\rangle$ and $|f\rangle$ states causing spatial shift towards the Mo atoms in real space is driven by the geometric Berry connection in the energy space.[17]

To analyze the complete temperature-dependence of the overall THz signal (Fig. 1(c)), we propose an extended Varshni's model, comprising the contributions from the excitonic shift current, the TPE, and the OR effects, given as

$$E_{THz}^{PP}(T) = \left(E_{THz}^{PP}(0) - \frac{\alpha T^2}{(\beta+T)}\right) - \gamma T + \delta \tag{1}$$

Here, $\alpha$, $\beta$, $\gamma$, and $\delta$ are real positive constants. The first term on the right-hand side (inside small brackets) is motivated from Varshni's model[33] for the exciton binding energy and hence accounts for the excitonic contribution. Here, $E_{THz}^{PP}(0)$ represents the maximum value of the THz radiation attainable from the $J_{ExS}$ at the lowest temperature while, the constant $\beta$ determines the Debye temperature of the system. The second term ($\gamma T$) is for the contribution by TPE in which an increased scattering of the photocarriers with the increasing temperature reduces the carrier mobility[34] and hence the $J_D$[23]. The constant $\delta$ accounts for a near temperature-independent contribution to THz emission by OR[35, 36]. The sum of all three components, according to Eq. (1), is plotted in Fig. 1(c) where a good fit (solid curve) with the data validates the consistency of the model used here, particularly, the ultrafast excitonic shift current as derived from Varshni's temperature-dependent exciton binding energy model. The fit yields $\beta$ ~$260\ K$, i.e., a value consistent with the Debye temperature[37] of the bulk $MoS_2$.

The excitation fluence ($F$) dependence of the $E_{THz}^{PP}$ was measured at a few specific temperatures, as shown in Fig. 2 for $300\ K$, $200\ K$, and $150\ K$, and in Fig. 3 for 20 K. These results also support our initial proposition that the excitonic shift current is the major contributor to the THz emission at low temperatures. From Figs. 2 and 3, we note that, except for the peculiar behavior at $20\ K$, a strong saturating behavior of the $E_{THz}^{PP}$ with the excitation fluence is present which makes it apparently clear that the THz radiation generation via OR is relatively much weaker. At $300\ K$, the amount of THz emission via TPE increases quickly with excitation fluence up until a saturation value $F_S$. Above $F_S$, the excessive photocarriers incoherently scatter from the phonons more than they can produce a coherent photocurrent $J_D$. At the lower temperature of $250\ K$, there is a substantial fraction of photogenerated excitons that constitute $J_{ExS}$ and hence take part in the THz radiation generation concurrently with the photocarriers. As we lower the temperature further, the density of photogenerated excitons increases much rapidly with the increasing $F$, and so is its contribution to the THz emission via coherent $J_{ExS}$. The saturation in the THz emission is attained much quicker at lower sample temperatures, and the difference in the magnitudes is indicative of the role of the $J_{ExS}$.

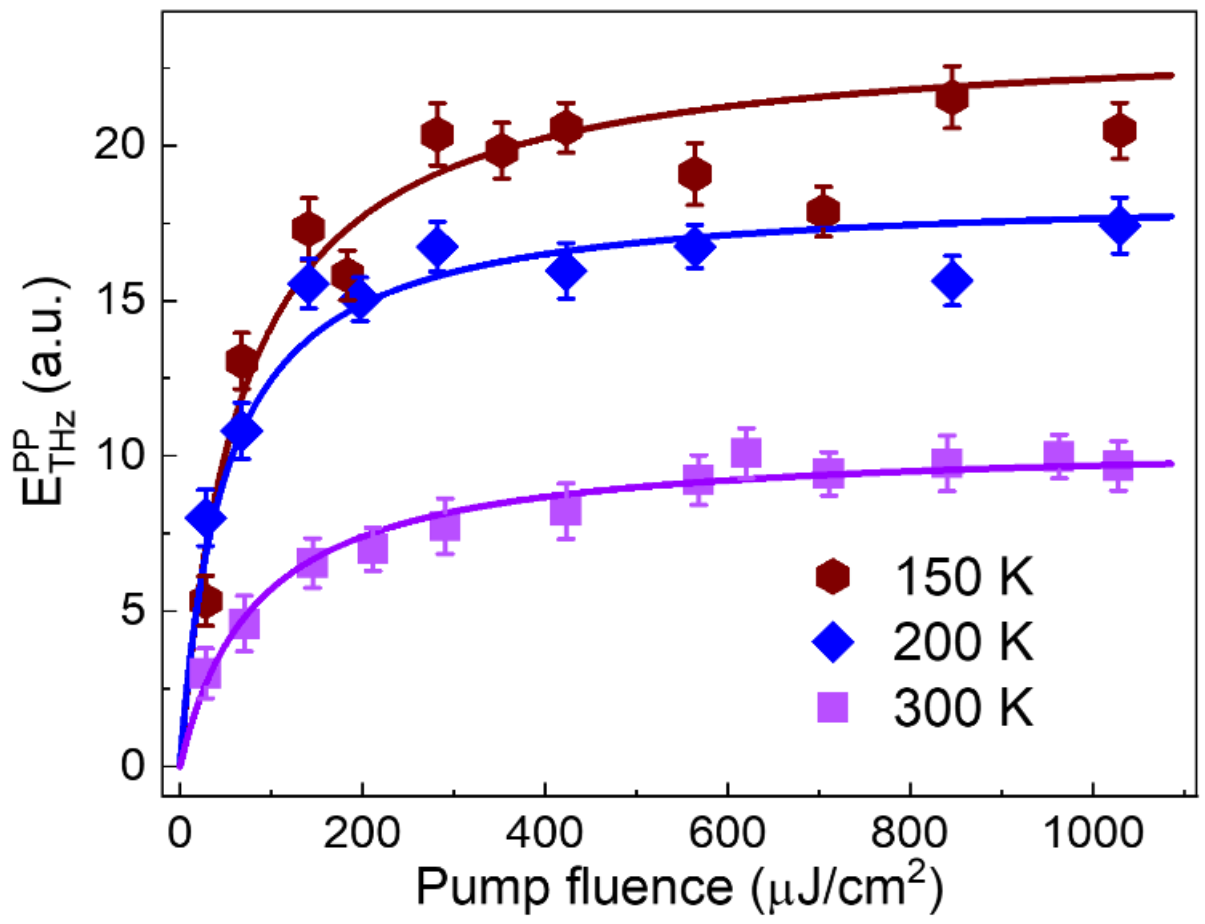

**Figure 2. Photoexcitation fluence-dependent THz emission at various sample temperatures. The solid curves represent the model given in Eq. (2) for extracting the saturation fluence ($F_S$).**

At a fixed sample temperature, we have used a phenomenological model given by[14]

$$E_{THz}^{PP} = S_1 F + S_2 \left[1 - \frac{1}{(1+F/F_S)}\right] \tag{2}$$

to capture the saturating behavior of the $E_{THz}^{PP}$ with the excitation fluence $F$. Here, $S_1$ and $S_2$ are constant scaling factors. The first term of Eq. (2) quantifies the contribution to THz emission due to the nonlinear OR and $J_{ExS}$, while the second term shows the contribution of saturating $J_D$. The obtained values of $F_s$ exhibit an increasing trend with the temperatures, but they remain well below $120\ \mu J/cm^2$. The saturation fluence values, along with those of the constant scaling factors $S_1$ and $S_2$ as obtained from the fitting at different sample temperatures, are provided in S5 of the Supplementary information.

As can be seen from Fig. 3, at the lowest temperature of $20\ K$, the exciton density and hence the amount of THz generation due to $J_{ExS}$ rises rapidly with the increasing fluence. Beyond a critical excitation fluence $F_C \sim 150\ \mu J/cm^2$, a sudden drop in the $E_{THz}^{PP}$ magnitude is seen, which then attains a much lower and nearly constant value at higher fluences. A less prominent drop in the THz generation efficiency near the critical fluence is also evident from the result at 150 K (Fig. 2), where the stable exciton density is not significantly high and hence the excitonic shift current is not yet the dominant one toward THz emission. Above the critical fluence, the sudden drop in the THz emission due to excitonic shift current is indicative of a critical density of the photogenerated excitons at which the free exciton-rich state transitions to an electron-hole liquid condensate at low temperatures.

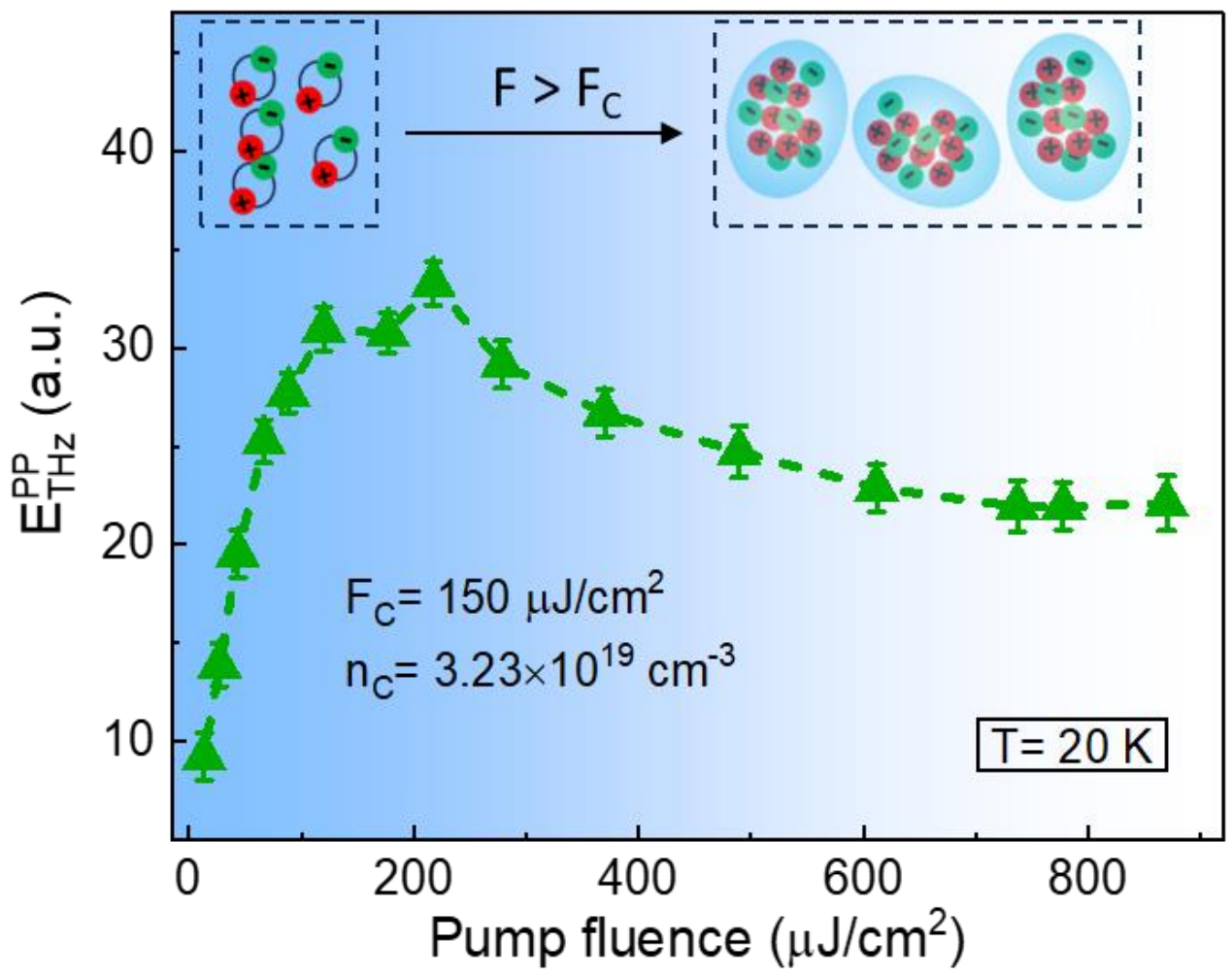

**Figure 3. Photoexcitation fluence-dependent THz emission from bulk $MoS_2$ at the lowest sample temperature of $20\ K$. Insect represents a rapid formation of excitons with the increasing fluence and transition to electron-hole liquid condensate beyond a critical excitation fluence $F_c$~$150\ \mu J/cm^2$.**

Two regions of the excitation fluence, i.e., $F < F_c$, and $F > F_c$, are distinguished in Fig. 3. When $F < F_c$, the exciton density increases rapidly with the excitation fluence, and hence an enhanced THz emission from the excitonic shift current follows up. At $F = F_c$, the exciton density is so high that the average spacing between exciton pairs is nearly the same as the exciton radius. Therefore, the pairwise Coulomb interaction between electrons and holes of the excitons screens out for them to rather be dissociated into separate electrons and holes ensemble[38]. This situation rather favors the formation of an electron-hole liquid condensate[39, 40]. Moreover, at high fluences, bandgap renormalization[7] causes a reduction in bandgap energy and the exciton binding energy. Hence, in the $F > F_c$ region, due to the reduced binding energy, the remaining excitons are also dissociated into charge carriers leading to formation of many body condensate, i.e., electron-hole liquid droplets[39, 40]. Within the self-sustained and confined volume of this new macroscopic quantum state, the electrons and holes move independently from one another. The decrement in the THz emission by drop in exciton density above $F_c$ persists until $F \sim 580\ \mu J/cm^2$ at which the exciton density becomes nearly null and only the $J_D$ contribute at its saturation limit as indicated in Fig. 3.

Not only does the critical density of excitons ($n_c$) corresponding to critical fluence, but also a temperature lower than a certain critical temperature ($T_c$) has to be appropriate so as to favor the transition from the free exciton-state to the correlated EHL state. From our data, we now estimate the values of $n_c$ and $T_c$ for the formation of the EHL macrostate. Since, the critical temperature is related to the exciton binding energy[38] $E_{exb}$ as $k_b T_c < 0.1 E_{exb}$, where $k_b$ is Boltzmann constant, using $E_{exb} \sim 50\ meV$ for bulk $MoS_2$,[5] we obtain $T_c < 60\ K$. Considering the quasi-equilibrium condition for the excitons, the critical density of the excitons is related to the critical fluence by the relation,[38] $\frac{aF_c}{\tau_p E_p} = \frac{n_c}{\tau}$, where $a$ is the absorption coefficient at excitation photon energy $E_p$, $\tau_p$ is the excitation pulse duration and $\tau$ is the stable exciton mean life-time. We obtain a value, $n_c \sim 3.2 \times 10^{19}\ cm^{-3}$ and $F_c = 150\ \mu J/cm^2$ for $MoS_2$ at $E_p = 1.55\ eV$. This value is very close to the Mott density in bulk $MoS_2$. Please see Section S6 for further details.

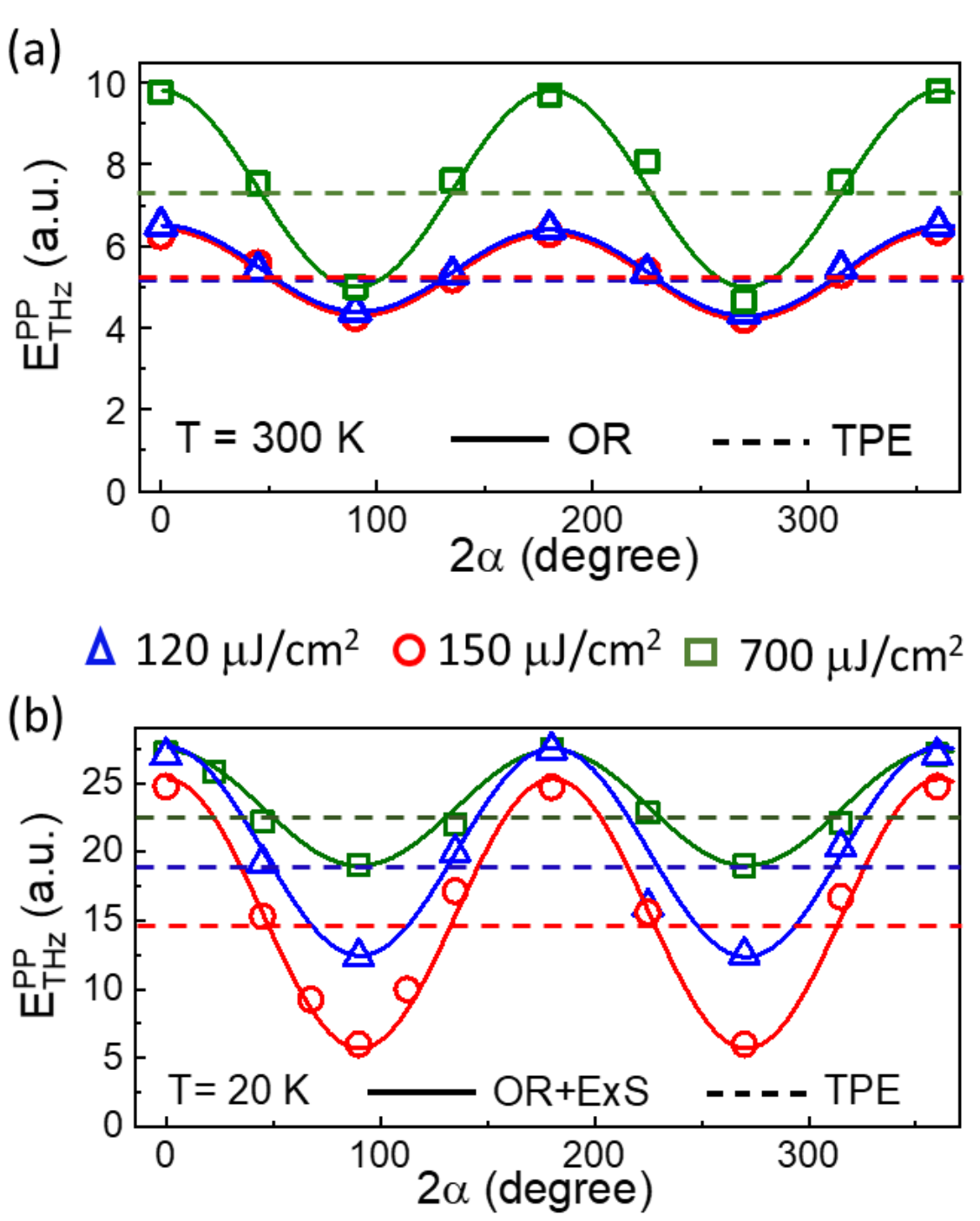

**Figure 4. Variation of THz generation efficiency with respect to the excitation polarization angle ($\alpha$) obtained at (a) 300 $K$ and (b) 20 $K$ under different excitation fluence values of 120 $\mu J/cm^2$ ($F < F_c$), 150 $\mu J/cm^2$ ($F \sim F_c$), and 700 $\mu J/cm^2$ ($F > F_c$).**

Results obtained from the pump excitation pulse polarization angle ($\alpha$) dependent experiments are summarized in Fig. 4 for the sample at the two extreme temperatures of 300 $K$ and 20 $K$. These were taken at fixed values of the fluences, $F \sim F_C$ (150 $\mu J/cm^2$), $F < F_c$ and $F > F_c$. The experimental data in Fig. 4, represented by symbols, show a two-fold symmetric variation that is riding over a constant offset. This behavior can be well captured by the standard polarization-dependent model relation[14], $E_{THz}^{PP} = A\, cos(2\alpha) + B$, as shown by solid lines. At 300 $K$ temperature, the coefficients $A$ and $B$ are related to the OR and surface field-induced TPE contribution, while at 20 $K$, these coefficients are related to the OR+ExS and TPE processes, respectively. At 300 $K$, the contribution of the TPE process increases as the excitation fluence increases, specifically, at 700 $\mu J/cm^2$, the TPE process is more dominant than the OR effect. On the other hand, at 20 $K$ temperature and at critical pump fluence, polarization dependence is primarily due to the dominating excitonic shift current. These observations are consistent with the results presented earlier in Figs. 2 and 3 in the paper.

To summarize, we have investigated the excitonic shift current mediated THz emission from bulk $MoS_2$ upon photoexcitation by broadband femtosecond laser pulses centered at 800 nm. The experiments were carried out at varying temperatures, the excitation fluence, and the polarization, which helped us to extract the relative contribution of the excitonic shift current over the usual transient photocurrent towards the THz generation. More importantly, at low temperatures where the excitons are the major contributor to the THz emission, we have observed a peculiar behavior in the fluence dependence of the THz emission, where the transition from a free excitonic state to a correlated EHL macrostate can be attributed beyond a critical density of the excitons. Clearly, by analyzing the THz emission from material systems, time-domain THz emission spectroscopy offers a novel tool for understanding photoinduced transitions noninvasively.

# Methods

# Excitonic shift current induced broadband THz pulse emission efficiency of layered single crystalline $MoS_2$

Neetesh Dhakar and Sunil Kumar*

*Femtosecond Spectroscopy and Nonlinear Photonics Laboratory, Department of Physics,*

*Indian Institute of Technology Delhi, Hauz Khas, New Delhi 110016, India*

**Corresponding author: kumarsunil@physics.iitd.ac.in*

The THz emission measurements were conducted using a home-built THz time-domain spectrometer, schematically presented in Fig. 5. This experimental setup is highly versatile, allowing operation at both room temperature[1-3], and low temperatures[4-6]. For temperature dependent measurements, we incorporated a liquid helium optical cryostat in the experimental setup. The cryostat chamber is designed with specialized windows to facilitate the passage of the optical and THz beams. The setup can be pumped by output from a near-infrared (NIR) laser source or a parametric amplifier, where the THz pulses are detected using a gating beam taken from the NIR laser source. In the current study, a regenerative Ti:S amplifier at 800 nm and a parametric amplifier delivering pulses at different wavelengths were used for the sample excitation. The pulse duration was ~35 fs, and the repetition rate of 1 kHz. For driving the setup with the NIR pulses from the amplifier, the laser beam is initially divided into two parts using an optical beam splitter (BS), where the stronger portion serves as the pump beam and the much weaker portion acts as the probe (gating pulse). To enhance the signal-to-noise ratio in THz pulse detection, the pump beam is modulated using an optical chopper operated at ~267 Hz. THz radiation is collected in the reflection configuration by using a pair of off-axis parabolic mirrors. The collimated THz beam is subsequently focused onto a 500 μm thick, [110]-oriented ZnTe nonlinear crystal for detection. Simultaneously, a low-power probe beam, which travels through a mechanical delay stage, propagates collinearly with the THz beam towards the nonlinear crystal. A quarter-wave plate (QWP), a Wollaston prism (WP), a balanced photodiode (PD), and a lock-in amplifier help to detect the THz pulse by electro-optic sampling, where the differential signal output from the lock-in amplifier directly provides a measurement of the THz pulse's instantaneous electric field as a function of time. By continuously varying the probe pulse delay relative to the THz pulse, the temporal profile of the THz signal is recorded. The mathematical overview of the whole process is described below.

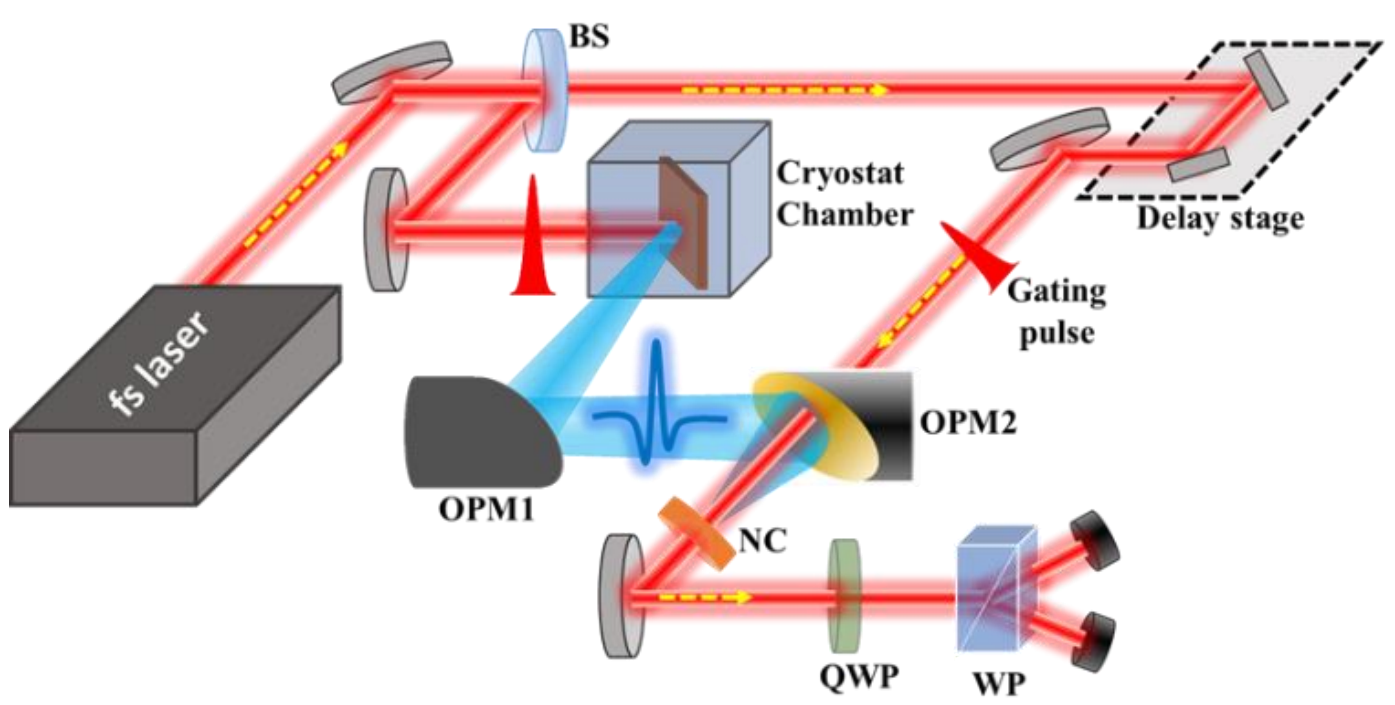

**Figure 5. Schematic of time-domain THz spectroscopic experimental setup. Pulsed THz generation from the sample after photoexcitation by a femtosecond NIR pulse and probing of the THz pulses by electro-optic sampling in a nonlinear optical crystal. BS: beam splitter, OPM1, and OPM2: off-axis parabolic mirrors, QWP: quarter-wave plate, NC: nonlinear optical crystal, WP: Wollaston prism, PD1, and PD2: photodiodes.**

A time-delayed gating pulse, $G(t-\tau)$ samples the THz field, $E_{THz}(t)$ at each of the controllable time delay, $\tau$ such that a convolution function, $S(\tau)$ builds up as a function of $\tau$ and given by the expression, $S(\tau) = \int_{-\infty}^{+\infty} E_{THz}(t)\, G(t-\tau)dt$. Since the gating pulse is much shorter, it can be assumed to be a Delta function, and hence the Fourier transform shift theorem is applied to obtain, $E_{THz}(\tau) = \int_{-\infty}^{+\infty} E_{THz}(t)\, \delta(t-\tau)dt$ or $E_{THz}(t) = \int_{-\infty}^{+\infty} E_{THz}(t')\, \delta(t'-t)dt'$. Thus, the THz pulse is detected in real time. In the balanced detection setup, the instantaneous birefringence induced in the nonlinear crystal by the strength of the THz pulse field at every time delay rotates the linear polarization of the gating pulse. This is simply the linear electro-optic effect, where the induced nonlinear material polarization is $P(t) = \int_{-\infty}^{+\infty} \varepsilon_0 \chi^{(2)}(t,t') E_{op}(t') E_{THz}(t') dt' \sim \sum_k P(\omega_k) e^{-i\omega_k t} \sim \sum_k \varepsilon_0 \chi^{(2)}(\omega_k; \omega_{op}, \pm\Omega_{THz}) E_{op}(\omega_{op}) E_{THz}(\pm\Omega_{THz}) e^{i\{(k_{op} \pm k_{THz}).\bar{r} - \omega_k t\}}$. Since, $\Omega_{THz}$ is negligibly small in comparison to the optical gating pulse, the source $P(t)$ emits an optical field of nearly the same frequency (within the broadband spectrum of the short gating pulse) but with rotated polarization. Thus, in the balanced detection, the difference in intensity recorded in the two photodiodes is exactly proportional to the magnitude of the THz pulse field at that instant. The whole of the above process is described through schematics in the figure below. It may be noted that we get only the proportional amount of the THz signal, whose actual strength can be measured by using a proper calibration as discussed in the previous literature, including a report from our group[2].

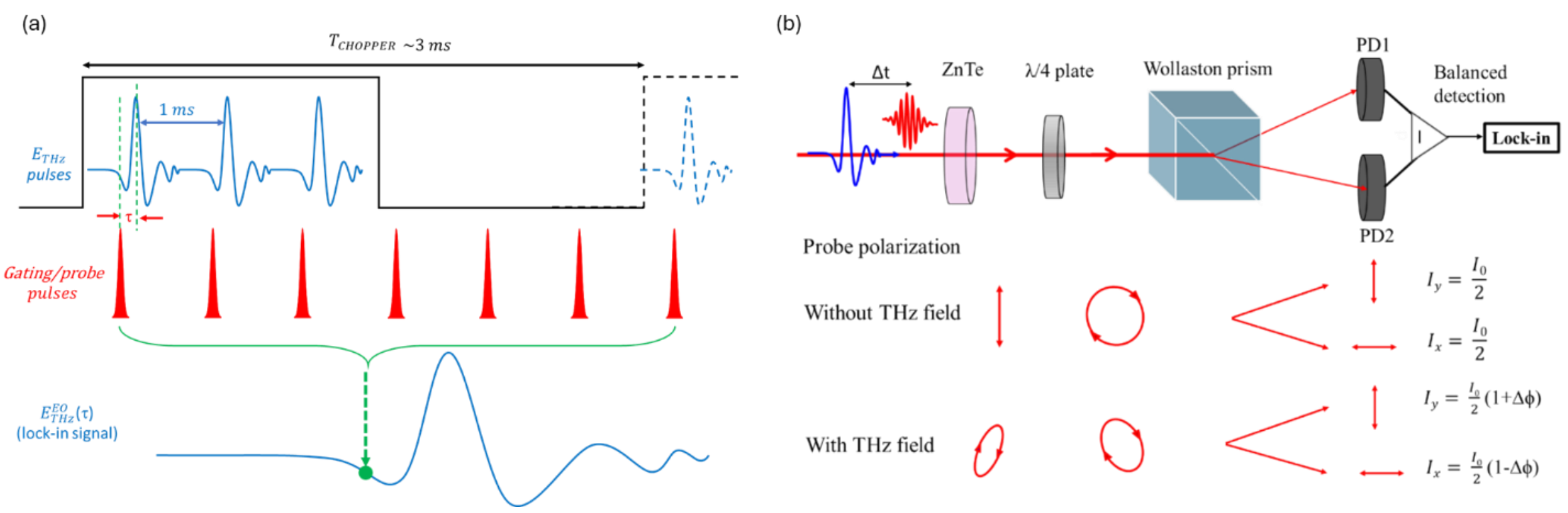

**Figure 6. An schematic representation of the process of THz pulse detection through electro-optic sampling technique.**

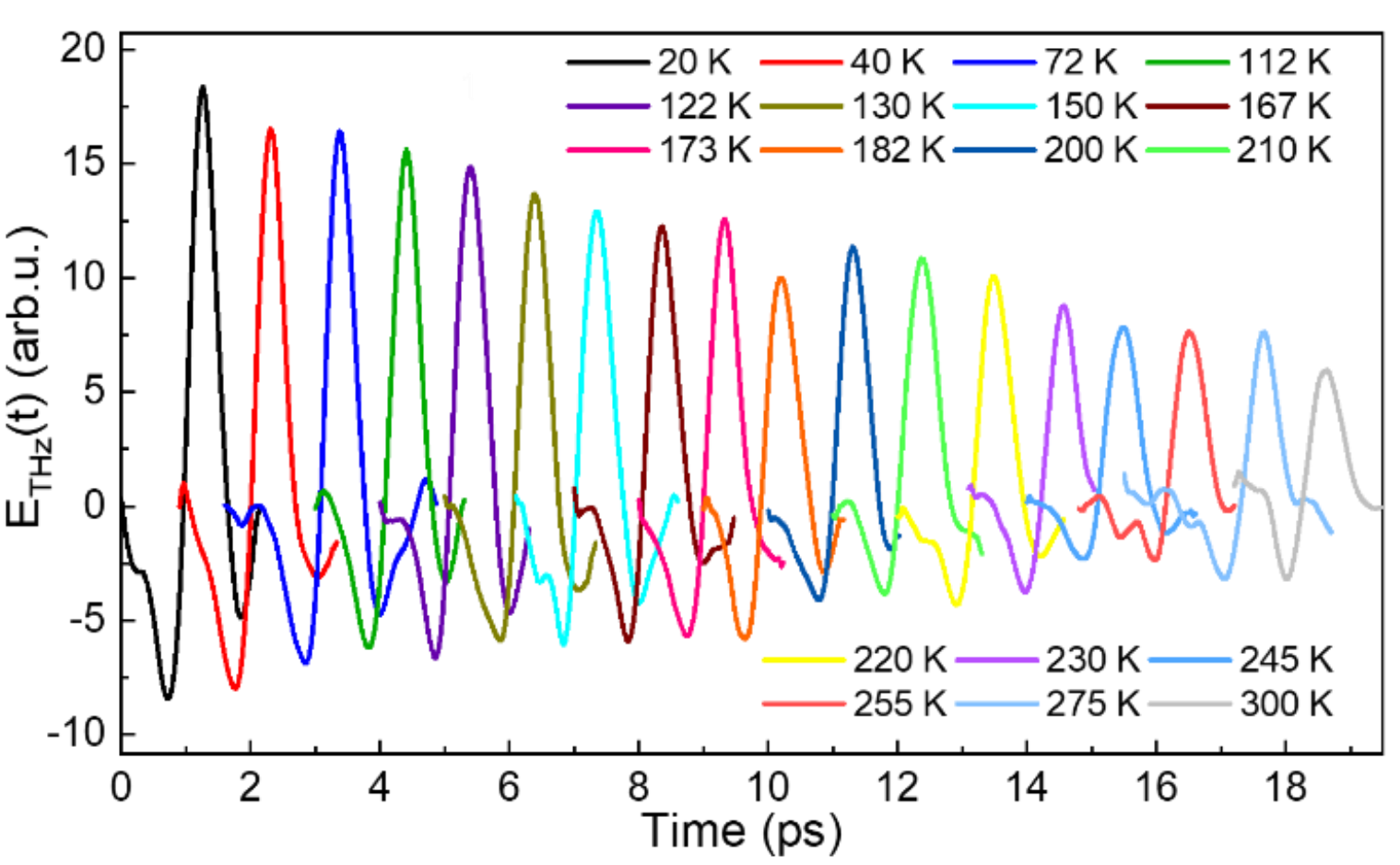

**Figure 7. THz Time-domain waveforms as obtained from the ultrafast photoexcited bulk $MoS_2$ at varying sample temperatures as indicated. The scans are relatively offset horizontally for better clarity.**

To identify the phase of the $MoS_2$ crystal in the current study, X-ray diffraction (XRD) measurements were performed. The corresponding results are presented below in Fig. 8. The characteristic diffraction peaks at $2\theta$ = 14.74°, 29.14°, 44.38°, 60.38°, and 77.78° correspond to the (003), (006), (009), (0012), and (0015) planes, respectively, and hence confirm the layered nature of the crystal. A magnified view is also presented in the figure which helps to observe the low-intensity peaks at $2\theta$ = 39.74°, 52.09°, 54.99°, 69.15°, and 87.32°, identified as diffraction from the planes (104), (018), (110), (108) and (118), respectively, prisence of which is consistent with the consistent with the literature[7, 8] for 3R phase of the $MoS_2$ crystal under study.

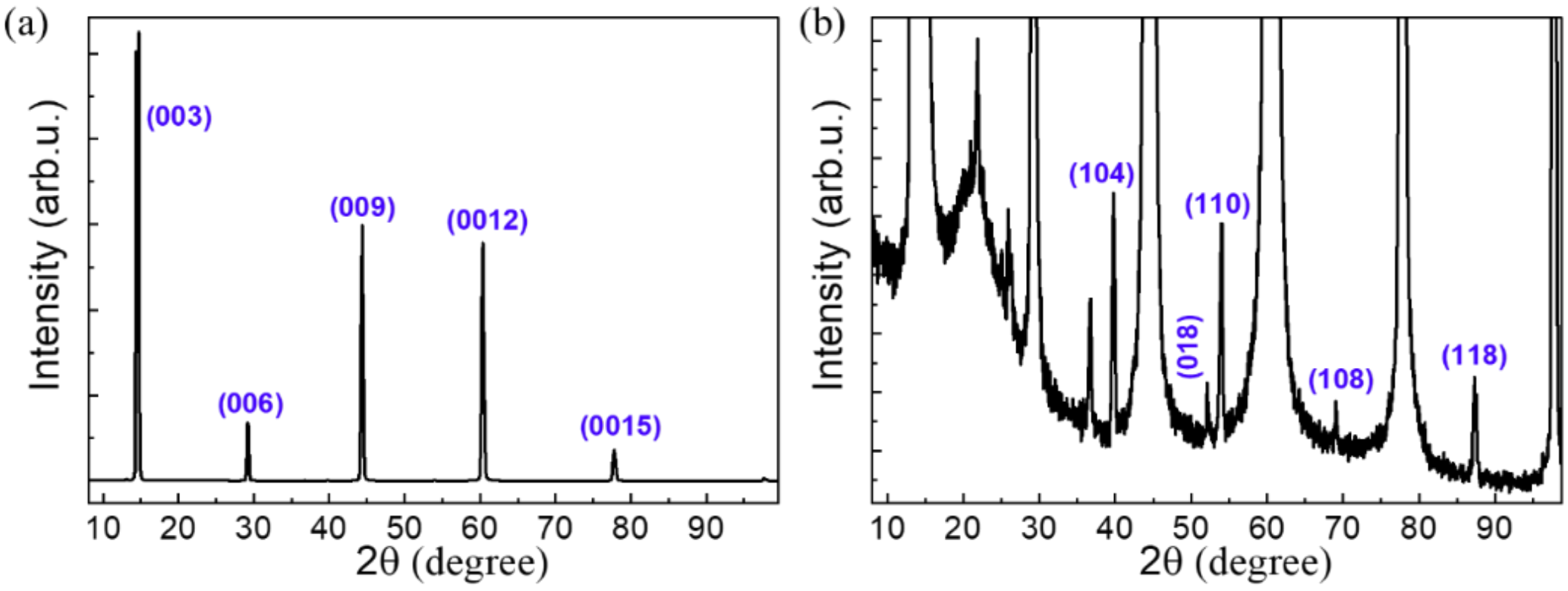

**Figure 8. Crystal phase identification from the XRD pattern. (a) Experimentally obtained data, and (b) the zoomed-in view for identifying the weak diffraction peaks that are characteristic of the 3R phase of the $MoS_2$ single crystal used in the current study.**

The extended form of the Varshni model (given by Eq. 1 in the manuscript) is appropriate to reproduce the behavior of the THz emission in the entire experimental temperature range as presented in Fig. 9. The values of the different parameters obtained from fitting are: $\alpha$ =12.8 units; $\beta$ = 260 K; $\gamma$ = 0.5 units; δ = 200 units; and the individual contributions are also indicated by dashed curves in Fig. 9.

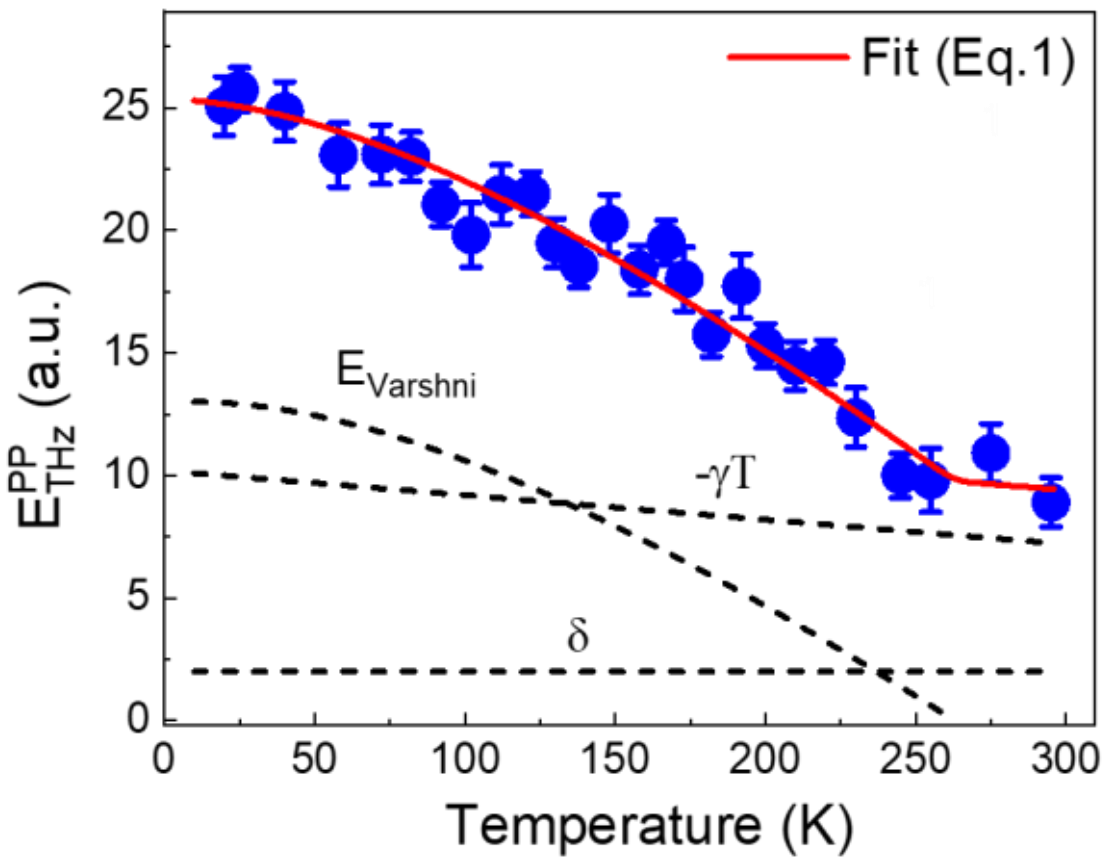

**Figure 9. Temperature dependent THz generation efficiency of the bulk $MoS_2$ single crystal. The red continuous curve represents modal fit using the extended Varshni model, while the dashed curves represent the individual components, i.e., the exciton effect via the Varshni model, the transient photocurrent effect (-$\gamma T$), and the optical rectification (δ).**

At higher temperatures, excitons become thermally unstable and dissociate into free electrons and holes, leading to an increase in free carrier density, which directly impacts THz emission by the exciton shift current as well as the transient photocurrent. The term $-\gamma T$ represents the contribution of surface depletion field-driven photocarriers, which induce coherent photocurrent, where the resulting drift photocurrent primarily depends on carrier mobility ($\mu$) and follows the relation[9] $J \propto N(F,T)\,\mu(T)$. Here, $N(F,T)$ is the photocarrier concentration at the given pump fluence $F$ and sample temperature $T$. Because the photogeneration takes place across the indirect bandgap, hence, $N(F,T)$ can be assumed to be weakly increasing with the temperature, along with the exponentially growing population of thermally generated incoherent carrier density at high temperatures. Owing to the decreases in the carrier mobility in $MoS_2$ at both the low and high temperatures[10, 11], the cumulative effect of both the above to the THz generation can be safely approximated by $-\gamma T$, where $\gamma$ is the appropriate scaling factor. Furthermore, since the second-order nonlinear susceptibility, $\chi^{(2)}$, is relatively nearly temperature independent[12, 13], hence the corresponding THz radiation generation by the optical rectification in the $MoS_2$ crystal upon ultrafast photoexcitation at an oblique incidence is finite but can be considered to be temperature independent.

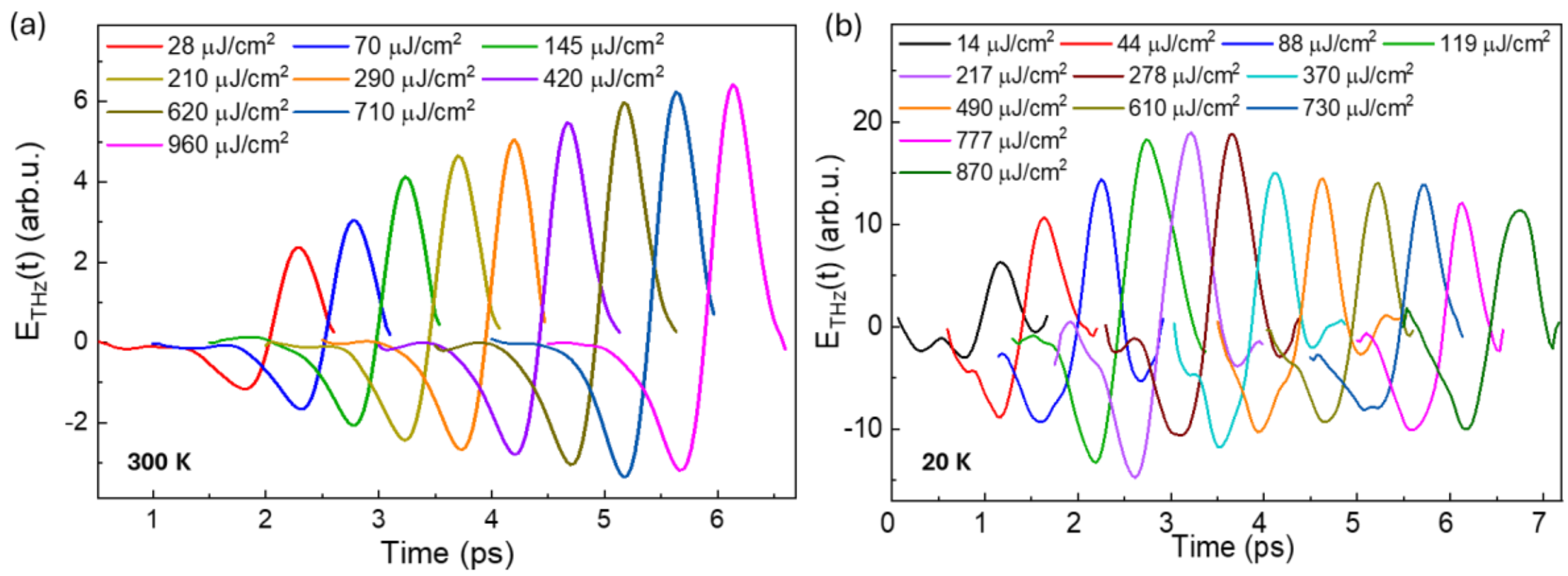

**Figure 10. Time-domain THz waveforms recorded at different values of the excitation fluence as indicated for the sample temperature of (a) 300 K, and (b) 20 K. The THz scans are relatively offset horizontally for clarity.**

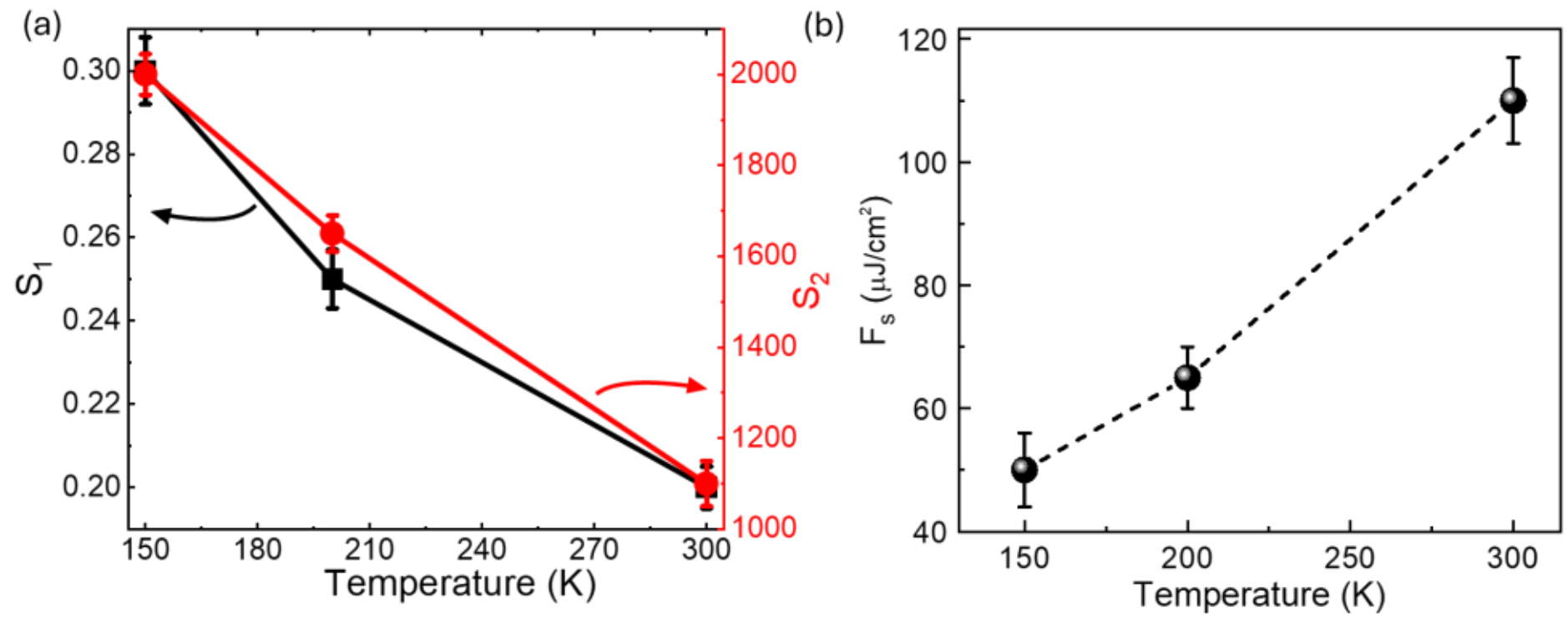

**Figure 11. (a) Values of the constant scaling factors $S_1$ and $S_2$, and (b) saturation fluence ($F_s$) at different temperatures as obtained by fitting the data in Fig. 2 using the phenomenological model given by Eq. (2) in the manuscript.**

The Mott density ($n_m$) refers to the concentration at which excitons ionize into free electrons and holes and possibly transition to the electron-hole liquid (EHL) phase beyond that. Typically, the Bohr radius is $a_B \sim 1 - 3\,nm$ in bulk $MoS_2$[14]. Taking $a_B = 2\,nm$, the Mott density comes out to be $n_m = \frac{3}{4\pi a_B^3} \sim 3 \times 10^{19}\,cm^{-3}$. The transition from free-exciton phase to the correlated EHL phase is taking place beyond the optical excitation fluence of $F_c \sim 150\ \mu J/cm^2$ corresponding to which the exciton density ($n$) is above the Mott density for excitons. The exciton density is related to the pump fluence $F$ via the relation[15], $\frac{aF_c}{\tau_p E_p} = \frac{n}{\tau}$, where, $a$ is the absorption coefficient at excitation photon energy $E_p$, $\tau_p$ is the excitation pulse duration, $\tau$ is the stable exciton mean-lifetime. After plugging in the experimental values of $F_c$ = 150 μJ/cm$^2$, $E_p$= 1.55 eV, $\tau_p$= 35 fs and taking the values of $\tau$ and $a$ from the literature[16, 17] as $\tau$ = 1 ps and $a$ = 1.87×10$^3$ cm$^{-1}$, respectively, the critical exciton density of the phase transition is estimated to be $n_c = 3.23 \times 10^{19}\,cm^{-3}$. This value is quite close to the Mott density as described above.